# Evidence for Zinc Ion Sharing in Metallothionein Dimers Provided by Collision Induced Dissociation

Carlos Afonso*, Yetrib Hathout and Catherine Fenselau

Chemistry and Biochemistry Department, University of Maryland, College Park MD 20742, USA

*Corresponding author.

Present address: Université Pierre & Marie Curie, 4 place Jussieu, boite 45, 75252 Paris Cedex 05, France.

Tel: 33 1 44 27 32 64

Fax: 33 1 44 27 38 43

email : afonso@ccr.jussieu.fr

Pages: 11

Figures: 4

**Abstract**

Nanospray and collisionally induced dissociation are used to evaluate the presence and absence of interstrand co-chelation of zinc ions in dimers of two kinds of metallothionein. As was reported in a previous publication from this laboratory, co-chelation stabilizes the dimer to collisional activation, and facilitates asymmetrical zinc ion transfers during fragmentation. In the case of metallothionein, dimers of the holoprotein are found to share zinc ions, while dimers of metallothionein, in which one domain has been denatured, do not. Zinc ions are silent to most physicochemical probes, e.g., NMR and Mossbauer spectroscopies, and the capability of mass spectrometry to provide information on zinc complexes has widespread potential application in biochemistry.

In recent years electrospray ionization has been very useful for analysis of metal/protein complexes [1-4]. Applications to metalloproteins have been successful, largely because metal ions are usually complexed to proteins by strong electrostatic forces, and electrostatic forces are preserved and even reinforced in the water-free environment of the gas phase [5,6]. Electrospray and nanospray can provide information on identity and stoichiometry, e.g., the nature and number of metal ions present, with high mass accuracy and high sensitivity [7]. When relevant, information can also be obtained on the oxidation state of the metal ion [8,9]. More recently, collisionally induced dissociation (CID) has been used to provide information on cooperative binding within protein complexes, and on overall conformation [10,11]. We have previously proposed that CID can be used to distinguish zinc ion sharing or mixed ligand co-chelation in complexes of metallothionein and glutathione [4].

Metallothionein (MT) comprises two distinct domains, the N-terminal domain ($\beta$-domain) in which 9 cysteine residues coordinate three bivalent metal ions and the C-terminal domain ($\alpha$-domain) in which 11 cysteines coordinate four metal ions. Metallothionein functions mainly as a zinc ion reservoir, providing zinc ions to other zinc depleted metalloenzymes. Jiang et al [12] showed that each domain reacts differently. The $\beta$-domain is a better zinc donor than the $\alpha$-domain. It has also been demonstrated experimentally that metallothionein transfers zinc ions to other proteins, including proteins with lower zinc ion affinities, via intermolecular complexes[13-15]. It seems likely that zinc ions are transferred via rapid formation of intermolecular ligand fields [13,16].

Mass spectrometry is one of the few physical techniques that can provide information on biological zinc complexes, which are silent to most physicochemical probes [17]. Recently nanospray has been used to demonstrate that metal ions are rapidly redistributed between two molecules of MT, monitoring the scrambling that occurs after rapid mixing of $Zn_7MT$ with $Cd_7MT$ [17] Otvos et al. used NMR spectroscopy to provide evidence that this occurs via a transient dimeric intermediate [13]. No transient intermediate dimers could be detected by mass spectrometry. However, a stable dimeric complex of methallothionein as been characterized after isolation by size-exclusion chromatography, which appears to be stabilized by disulfide bridging [17].

The stoichiometry of this stable MT dimer in the gas phase is investigated here, and CID is used to address the question of interchain co-chelation of zinc ions. Control experiments are performed with a partially denatured MT dimer, to verify the specificity of the observed complexes and the validity of the CID method for characterizing "shared" zinc ions.

**Experimental**

Materials

Rabbit liver metallothionein 2a, zinc atomic absorption standard solution and ammonium bicarbonate were obtained from Sigma (St Louis, MO). Deionized water (18.2 MΩ) was obtained with a MilliQ (Millipore, Bedford, MA) apparatus. HPLC grade methanol and acetonitrile were obtained from Fisher Scientific (Pittsburgh, PA).

Sample preparation

Rabbit liver metallothionein 2a was purified by reverse phase HPLC and reconstituted with zinc ions, following a published procedure [14]. Briefly, metallothionein is denatured using hydrochloric acid and dipicolinic acid, and then the 2a isoform is purified by reverse-phase high pressure liquid chromatography (discarding the contaminating MT-IIb isoform). After lyophilization, the holoprotein $Zn_7MT$ is reconstituted by the addition of $ZnCl_2$, raising the pH slowly under argon to avoid oxidative dimerization. Excess zinc ions are eliminated by centrifugation using 5 kDa cutoff centrifugal ultrafiltration cartridge (Ultrafree, Millipore, Bedford, MA). Acetic acid (0.1 %) was added to this solution when partial denaturation was required.

An Applied Biosystems (Foster City, CA) Q-Star hybrid quadrupole/time of flight mass spectrometer equipped with a Protana (Odense, Denmark) nanoelectropray source was used for all the experiments. About 2 µl of sample was loaded into a Protana metallized tip using Eppendorf GELoaders 1-10µl (Westbury, NY). The instrument was set to work in the positive ion mode with an ionization voltage set at 900 V. The declustering potential was varied between 30 and 100 V. For tandem mass spectrometry, argon was used as collision gas and the collision energy was varied between 20 and 50 V. External calibration was performed using a polypropylene glycol 3000 solution (Applied Biosystems, Foster City, CA). The quadrupole mass filter was tuned to transmit ions in the 200-2500 m/z range. Simulation of isotopic distributions, and calculation of apoprotein monoisotopic and average masses were performed with an ICR-2LS v2.66.79 program (EMSL). Determination of protein formulae was performed using GPMAW 4.12 (Lighthouse data, Odense, Denmark).

**Results**

Under physiologic conditions (pH 7.4) in ammonium bicarbonate buffer, the ionized protein carrying 7 zinc ions was observed as expected (6569 Da in Figure 1). This mass is in good agreement with the theoretical average mass calculated from the amino acid sequence, plus seven zinc ions [7,18]. Excess zinc ions were eliminated before the analysis by ultrafiltration, and no nonspecific adducts are detected in the spectra. The narrow charge distribution observed here (centered around the 5+ ions) is typical for this stable folded protein [2]. Ions detected at m/z 1878 are assigned to the stable dimeric form of $Zn_7$-metallothionein carrying 7 protons. This dimeric species was found to be stable at relatively high declustering potentials (i.e. 100 V in Figure 1).

When the dimeric species carrying fourteen zinc ions was selected for collisional activation 50 V collision energy was required (Figure 2), and two complementary cluster ion distributions appeared, one with a +3 charge state and the other with a +4 charge state. The masses of each series correspond to the monomeric form of the protein with a continuum number of zinc ions. Although seven charges cannot be divided equally between the two CID product ions, the distribution observed indicates that the two polypeptides in the dimer have similar conformations and flexibility. The monomeric product ions with three charges carry from 7 to 12 zinc ions, with the 10 zinc specie as the most abundant. The monomeric protein product ions with four charges carry from 1 to 10 zinc ions. The 4 zinc specie is the most abundant.

This experiment was repeated with a solution of metallothionein that had been partially denatured at pH 3.5 by addition of diluted acetic acid (see Experimental). It has been shown previously that the two domains of MT have different stabilities [12]. Below pH 3 both domains are unfolded and all zinc ions are released. At pH 3.5 the β-domain unfolds and releases three zinc ions, whereas the α-domain remains folded, binding four zinc ions. Under this condition the mass spectrum displays mainly metallothionein carrying 4 zinc ions (presumably those cooperatively bound in the alpha-domain). When a low declustering potential (30 V in Figure 3) is used, a dimeric form is also observed, at m/z 1823, (Figure 3), carrying eight zinc ions and seven excess protons.

Under collisional dissociation (Figure 4), the dimer of partially denatured metallothionein is fragmented into MT monomers at two charge states (+4 and +3), each carrying four zinc ions. No evidence for any inter-cluster ion transfer was detected in the decompositon of this dimer.

Electrostatic interactions involving interstrand co-chelation of multiple zinc ions are expected to be relatively strong, and to stabilize the complex. It should be noted that higher CID energies were required to fragment the dimer of the fully metalated holoprotein, than were required to fragment the dimer of the partially denatured protein. A collision energy of 50 V was required to obtain the CID spectrum shown in Figure 2, while 30 V provided the CID spectrum in Figure 4.

**Discussion**

In an earlier study, CID was used to distinguish between $Zn_7MT$ complexes with glutathione derivatives that are stabilized by co-chelation of zinc ions, and those that are not [4], This distinction is based on differences in the collision energy required for dissociation, and more importantly, on differences in the distribution of zinc ions between the dissociation products. In the present study, differences are observed on both counts in the dissociation of the dimer of fully metalated MT and the dimer of the partially demetalated protein. The disulfide linkage postulated in the stable dimer of $Zn_7MT$ may contribute to the requirement for higher collision energies for dissociation. However, the asymmetric distribution of zinc ions in the products argues strongly for co-chelation of some or all between the two polypeptide strands. Reflecting on the distribution of 10 and 4 zinc ions in the most abundant dissociation products, one can speculate that the three metal ions in each of the labile beta-domains are involved in interstrand ligands to a larger extent than those in the more stable alpha domains, and thus are more readily transferred between polypeptides. This is also consistent with the absence of interstrand transfer between dimers of $Zn_4MT$. Co-chelation and transfer of metal ions from each beta domain would be symmetrically achieved in the head-to-tail conformation predicted for $Zn_7MT$ dimers based on the x-ray crystallographic structure [19] and supported by recent theoretical studies [20].

**Acknowledgment**

The authors thank Dr X. Yao and K. Reynolds for helpful discussions. This work was supported financially by the National Institutes for Health (GM 21248).

# Figure Captions

**Figure 1:** Mass spectrum of $Zn_7MT$ in a pH 7.4 ammonium bicarbonate solution.

**Figure 2:** Product ion spectrum of $[2Zn_7MT+7H]^{7+}$ ions from the spectrum in Figure 1, obtained with a collision energy of 50 V.

**Figure 3:** Mass spectra of $Zn_7MT$ partially denatured at pH 3.5.

**Figure 4:** Product ion spectrum of $[2Zn_4MT+7H]^{7+}$ from the spectrum in Figure 3, recorded with a collision energy of 30 V.

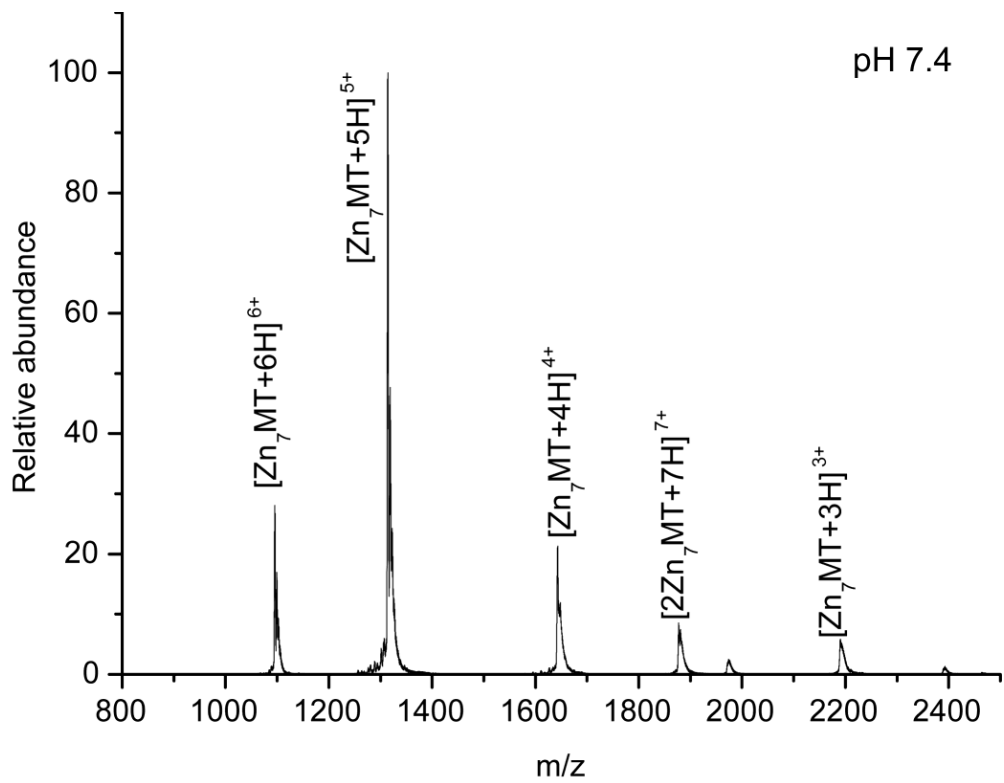

Figure 1

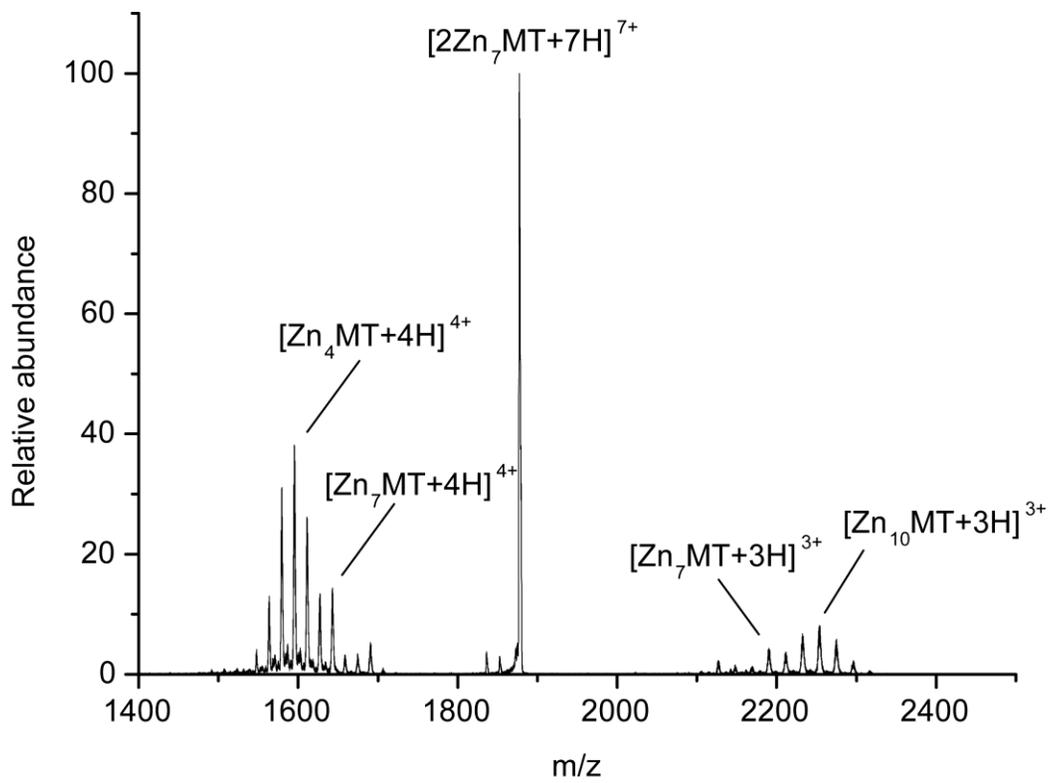

Figure 2

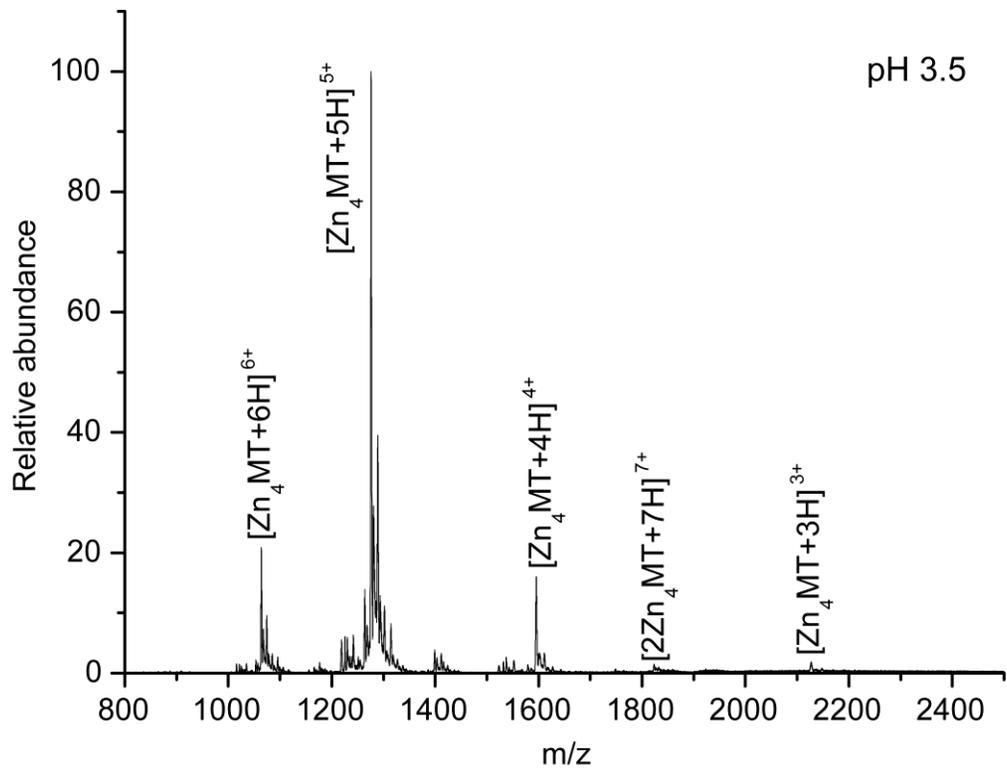

Figure 3

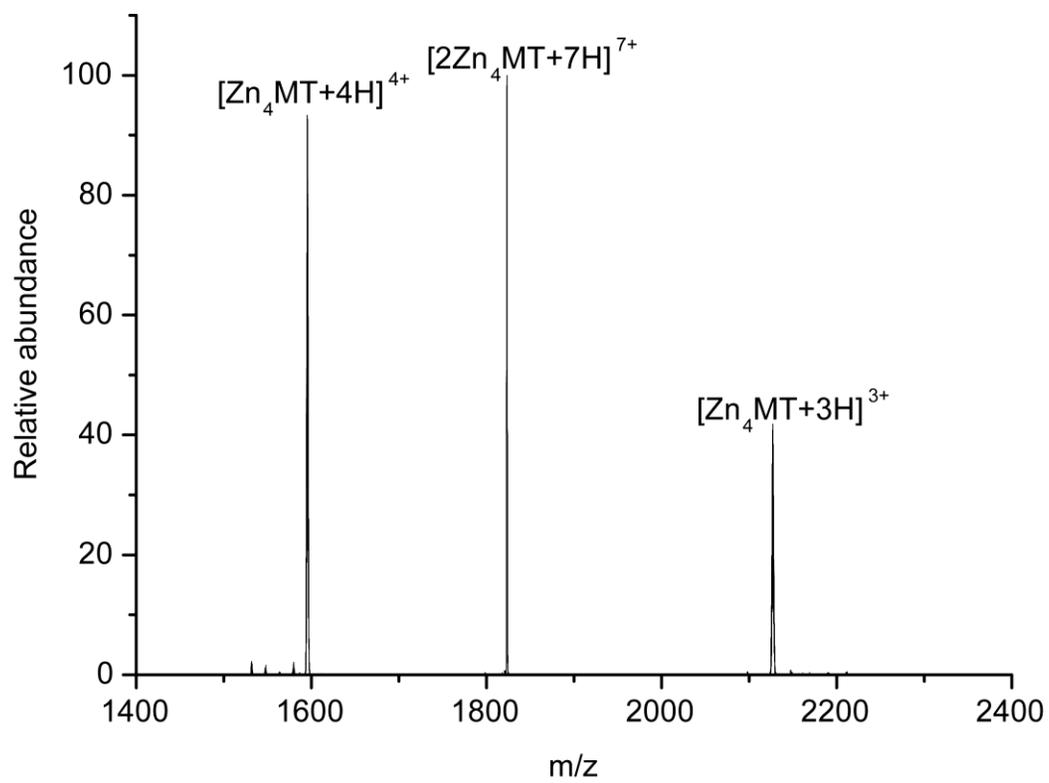

Figure 4